\documentclass[aps,prl,twocolumn,groupedaddress]{revtex4}
\usepackage{graphicx}
\usepackage{placeins}
\usepackage{float}
\usepackage{subfigure}
\usepackage{longtable}
\usepackage{amssymb}
\usepackage{filecontents}
\usepackage{natbib}

\begin{document}

\title{Unusual anisotropic magnetic orbital moment obtained from X-ray magnetic circular dichroism in a multiferroic oxide system}

\author{Daniele Preziosi}
\email[]{daniele.preziosi@ipcms.unistra.fr}
\affiliation{Université de Strasbourg, CNRS, IPCMS UMR 7504, 67034 Strasbourg, France}
\author{S. Homkar}
\affiliation{Université de Strasbourg, CNRS, IPCMS UMR 7504, 67034 Strasbourg, France}
\author{C. Lefevre}
\affiliation{Université de Strasbourg, CNRS, IPCMS UMR 7504, 67034 Strasbourg, France}
\author{M. Salluzzo}
\affiliation{CNR-SPIN Complesso di Monte S. Angelo, via Cinthia - 80126 - Napoli, Italy}
\author{N. Viart}
\affiliation{Université de Strasbourg, CNRS, IPCMS UMR 7504, 67034 Strasbourg, France}
\date{\today}

\begin{abstract}
The electric-field control of $d$-electron magnetism in multiferroic transition metal oxides is attracting widespread interest for the underlying fundamental physics and for next generation spintronic devices. Here, we report an extensive study of the $3d$ magnetism in magnetoelectric Ga$_{0.6}$Fe$_{1.4}$O$_3$ (GFO) epitaxial films by polarization dependent x-ray absorption spectroscopy. We found a non-zero integral of the x-ray magnetic circular dichroism, with a sign depending upon the relative orientation between the external magnetic field and the crystallographic axes. 
This finding translates in a sign-reversal between the average Fe magnetic orbital and spin moments. Large Fe-displacements, among inequivalent octahedral sites, lower the symmetry of the system producing anisotropic paths for the Fe-O bondings giving rise to a large orbital-lattice interaction akin to a preferential crystallographic direction for the uncompensated or averaged among different sites, magnetic orbital moment. The latter may lead to a partial re-orientation of the magnetic orbital moment under an external magnetic field that, combined to the ferrimagnetic nature of the GFO, can qualitatively explain the observed sign-reversal of the XMCD integral. The results suggest that a control over the local symmetry of the oxygen octahedra in transition metal oxides can offer a suitable leverage over the manipulation of the effective orbital and spin moments in magnetoelectric systems. 
\end{abstract}

\maketitle
\begin{figure}[t]
  \centering
\includegraphics[width=0.5\textwidth]{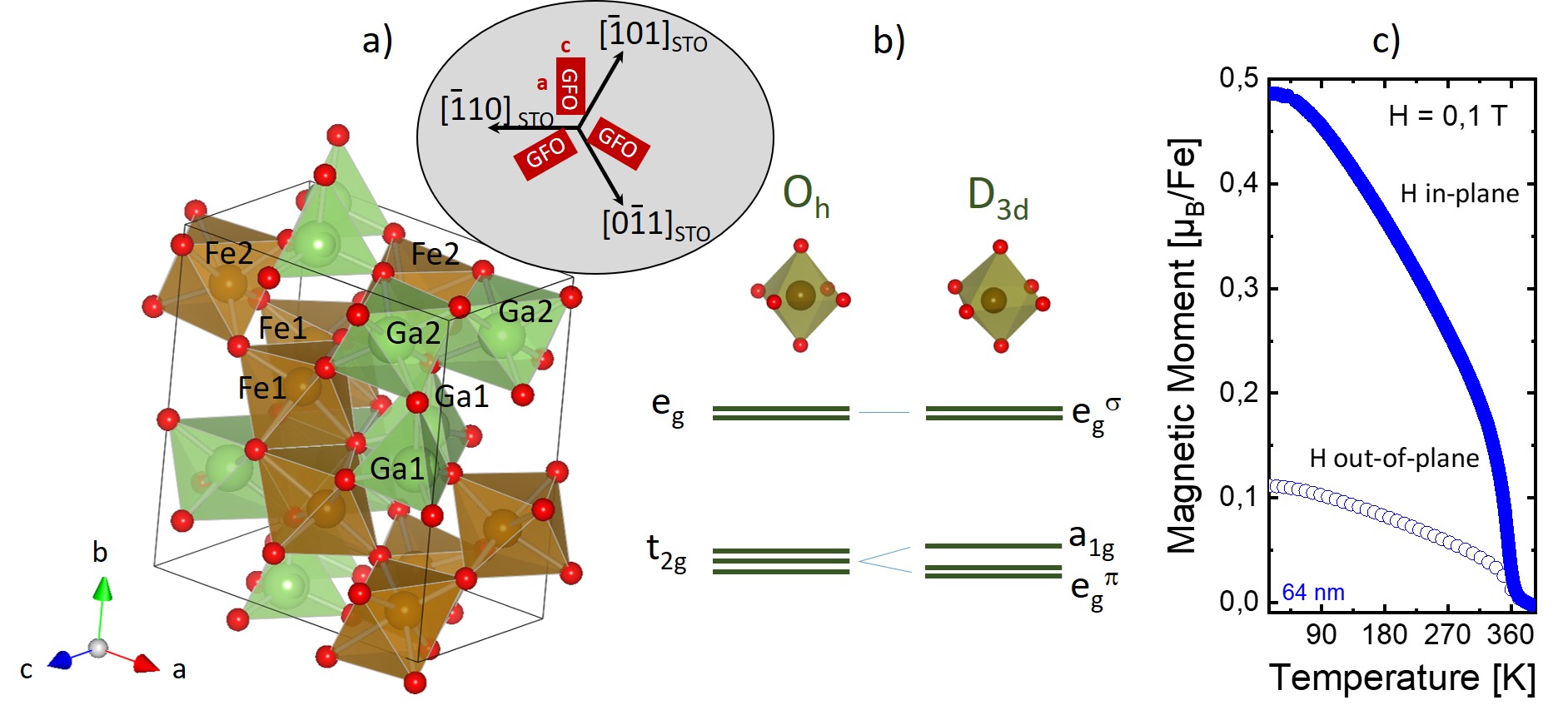}
\caption{\label{fig1} a) Conventional GFO unit cell with the four different sites. Inset shows the growth mode onto the STO(111). $\bf b$ also corresponds to the three-fold rotational axis of the b) trigonal distortion. c) Magnetization curves as a function of temperature for the in-plane and out-of-plane geometries. }
\end{figure}
\begin{figure} [t]
\includegraphics[width=0.45\textwidth, keepaspectratio]{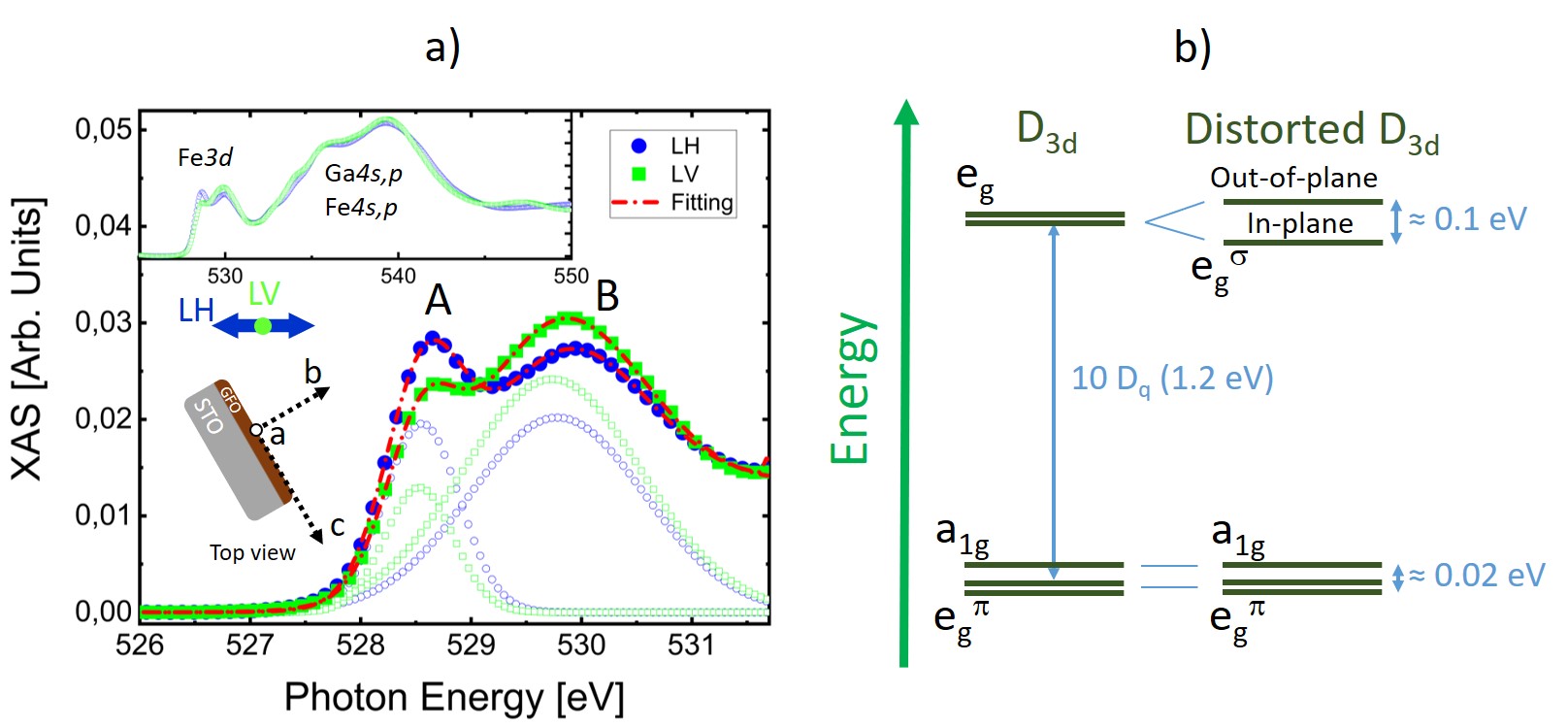}
\caption{\label{fig2} a) O-K edge XAS spectra acquired with LH and LV polarized light with a sketch of the geometrical setup. Inset shows the spectra for the entire energy range. The A- and B-labelled doublet (solid line) is reproduced (dashed line) by two Voigt functions (open-symbols). b) Derived energy diagram and energy splitting of the trigonal crystalline splitting with further degeneracy removal of the e$_g^{\sigma}$. }
\end{figure}
\begin{figure}
\includegraphics[width=0.42\textwidth, keepaspectratio]{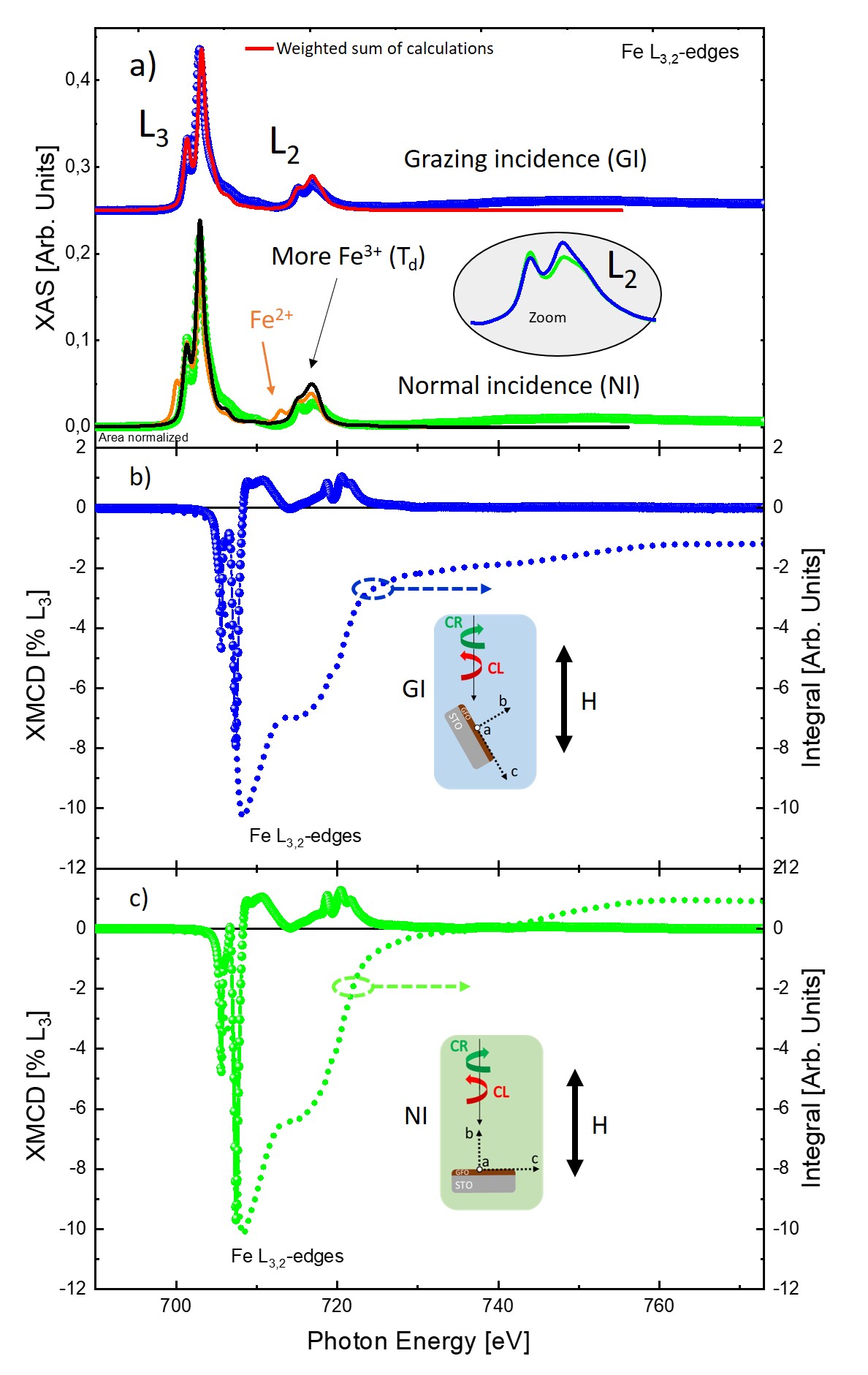}
\caption{\label{fig3} (a) XAS spectra corrected by a step function\cite{nakajima1999electron} acquired in GI  (blue circles) and NI geometries (green circles) compared to atomic multiplet simulation with parameters reported in Table 2. Red line is the simulation reproducing GI data.  Black and orange lines, overlapped to NI data, are the calculated spectra assuming a very large percentage of Fe$^{3+}$ in T$_d$ sites, or the presence of Fe$^{2+}$ ions in O$_h$ sites, respectively. The inset shows a zoom of the L$_2$ features. b) XMCD spectrum acquired at 6.5 T in GI (blue circles) and its integration on the whole range (blue dotted line), showing a negative integral value. c) XMCD spectrum acquired in NI at 6.5 T (green circles) and its integral curve (green dotted line), showing a positive integral value. The insets shows the experimental geometry. The XAS simulation was performed using a Gaussian broadening (experimental resolution) of 0.15 eV and Lorentzian broadenings of 0.2 and 0.4 eV for L$_3$ and L$_2$ features, respectively.}
\end{figure}

The emergent functionalities disclosed by transition metal oxides (TMOs) \cite{tokura2017emergent} together with the latest advances in atomic-scale synthesis \cite{boschker2011optimized}, are triggering a focus shift in the materials research panorama. The electric field control of the spin in ferromagnets (viz. spintronics), as envisaged in multiferroic TMOs \cite{banerjee2019oxide, vaz2018oxide}, is generating large interest for the potential high impact in technology. Multiferroics combine at least two ferroic order parameters (usually the ferroelectric polarization and the magnetization), which can conceptually lead to an efficient electric-field control of the d-electron magnetism. Recent propositions of devices based on the combination of multiferroic TMOs and large spin-orbit coupling (SOC) materials \cite{dowben2018towards, manipatruni2018beyond} would allow low dissipation writing and read-out processes within the same material stack, thus reducing the operational current in logic gates and memories, and enhancing the devices' scalability \cite{manipatruni2019scalable}. 
At the same time, recently it has been theoretically shown that some multiferroics, like BiCoO$_{3}$, can exhibit a bulk Rashba SOC \cite{PhysRevB.100.245115}. It follows that understanding how SOC couples the electron spin angular momentum to the electron orbital angular momentum in multiferroic TMOs is a necessary step towards a rational design of novel ultralow-switching-energy multiferroic-based devices.\\
Fe$^{3+}$-based oxides, being characterized by a $3d^5$ electronic configuration, should exhibit a zero magnetic orbital moment ($\mu_{L}$) as it is found, for example, in the case of $\gamma$-Fe$_2$O$_3$ nanoparticles \cite{BRICEPROFETA2005354}. 
However, a non-zero $\mu_{L}$, parallel to the magnetic spin moment ($\mu_{S}$), was reported in the case of magnetoelectric gallium ferrite (GaFeO$_3$) single crystals \cite{kim2006orbital} where, Fe$^{3+}$ ions occupy only octahedral (O$_h$) sites. This result was interpreted as a consequence of intrinsic FeO$_6$ crystalline distortions, capable of removing the parity symmetry of the $3d$-orbitals via anisotropic Fe$3d$-O$2p$ hybridization paths. Similarly, Tseng $et$ $al.$ 
by using the sum-rules reported a non-zero  $\mu_{L}$ parallel to $\mu_{S}$ in $\epsilon$-Fe$_2$O$_3$ nanoparticles \cite{tseng2009nonzero}. \\
In this work, we report polarization dependent X-ray absorption spectroscopy (XAS) measurements, and in particular Fe L$_{3,2}$-edges X-ray magnetic circular dichroism (XMCD) data, on magnetoelectric Ga$_{0.6}$Fe$_{1.4}$O$_{3}$(GFO) epitaxial films. The 0.4 excess of Fe increases the ferrimagnetic transition above room temperature. This Fe doping allows also a certain amount of Fe$^{3+}$ in tetrahedral (T$_d$) sites (see, for example, Table \ref{tab1}). 
By applying the sum-rules to the XMCD data we find that the magnetic orbital moment is anisotropic with a direction parallel or antiparallel to the net magnetic spin moment depending on the orientation between magnetic field and crystallographic axes. We explain the non-zero value of the net $\mu_L$ and the abnormal change of orientation with respect to the average $\mu_S$, as a consequence of a direction dependent compensation between $\mu_L$ and $\mu_S$ along different crystal axes of inequivalent (differently distorted) Fe sites. Here, the Fe ions moving away from the gravity center of the octahedral cage set a strong orbital-lattice interaction that, for large displacements may prevent the magnetic orbital moment to follow completely the externally applied magnetic field. \\
We studied 64 nm thick GFO films epitaxially grown onto (111) SrTiO$_3$ (STO) single crystals by pulsed laser deposition technique. Details on the growth and structural characterization are provided elsewhere \cite{homkar2019ultrathin}. The experiments were performed at DEIMOS beamline (SOLEIL Synchrotron facility - France) at 4 K in total electron yield mode (probing depth of $ca.$ 5 nm) with an energy resolution of $ca$. 0.15 eV \cite{Ohresser2014,Joly2014}. Each XMCD spectrum was obtained as the difference between an average of 8 XAS spectra acquired around the Fe L-edge, with a large pre- and post-edge energy range, and a magnetic field (H) parallel and antiparallel to the photon-helicity vector orientations. The 16 XAS data needed for each XMCD were collected in a sequence alternating reversal of field and circular polarization. This procedure ensures the best cancellation of spurious effects and experimental artifacts.
Gallium ferrite crystallizes in the orthorhombic $Pc2_1n$ (33) space group with four inequivalent cationic sites \cite{abrahams1965crystal}: tetrahedral Ga1 and octahedral Fe1, Fe2, and Ga2 sites. The ferromagnetically aligned spins at Ga1 and Fe1 sites are antiparallel to the ones at Ga2 and Fe2 sites, and the net magnetic moment depends on their relative occupations \cite{bertaut1966etude}. The conventional GFO unit cell is depicted in Figure \ref{fig1}a with $\bf b$- and $\bf c$-axis being the polar (ferroelectric) and magnetic (ferrimagnetic) easy axes, respectively.
\begin{table}[b]
\caption{\label{tab1} Cationic distributions of Fe$^{3+}$ and Ga$^{3+}$ over the four different cationic sites of the GFO orthorhombic structure. Experimental values (as taken from Ref. \cite{lefevre2016determination}), and theoretical ones as expected from the Ga$_{0.6}$Fe$_{1.4}$O$_3$ formula.}
\begin{ruledtabular}
\begin{tabular}{lccccccc}
&Ga1(T$_d$)&Ga2(O$_h$)&Fe1(O$_h$)&Fe2(O$_h$)&Sum&Expected\\
Ga$^{3+}$&0.60&0.25&0.15&0.05&1.10&1.20&\\
Fe$^{3+}$&0.40&0.75&0.85&0.95&2.95&2.80&\\
 \end{tabular}
 \end{ruledtabular}
 \end{table}
The anisotropic magnetoelectric properties of bulk GFO are largely associated to a magnetic field-induced modulation of the Fe displacements which occur at both Fe1 and Fe2 sites in opposite direction\cite{Arima2004a}. The Fe ions are assumed to re-adjust their relative position within the distorted FeO$_6$ octahedra mainly along the $\bf b$-axis when a magnetic field (H) is applied along the perpendicular direction ($i.e.$ the $\bf c$-axis). For the opposite situation, which sees H applied along the $\bf b$-axis, the Fe-displacements along the $\bf c$-axis are less energetically favorable and, most likely, controlled by the nature of the intrinsic distortion of the FeO$_6$ octahedra. The latter have one of the eight-faces oriented in the $\bf {ac}$-plane, and are trigonally distorted along the $\bf {b}$-axis which is also the three-fold rotational axis of the trigonal symmetry (D$_{3d}$)\cite{khomskii2014transition}. Here, the t$_{2g}$ Fe$^{3+}$ orbitals are transformed into a$_1^g$ and e$_g^{\pi}$ states, and the e$_{g}$ doublet becomes a unique e$_g^{\sigma}$ state, as schematically shown in Figure \ref{fig1}b. The growth orientation of the GFO onto the STO is sketched in the Inset of Figure \ref{fig1}a, where the bulk $\bf {ac}$-plane results stabilized on the substrate's surface and the bulk $\bf {b}$-axis coincides with the surface normal. \\
To attest the quality of our GFO thin films we performed magnetic measurements via vibrating sample magnetometry (Quantum Design, MPMS). Figure \ref{fig1}c shows the temperature dependence of the magnetization acquired with a magnetic field of 0.1 T oriented in both in-plane ($\bf {ac}$-plane) and out-of-plane ($\bf b$-axis) directions. While the Curie temperature of $ca.$ 356 K is the same for both geometries, the net magnetic moment is strongly anisotropic with the hard axis along the $\bf b$-axis (out-of-plane direction). The saturated magnetic moment measured at 10 K and 6.5 T is 0.78 $\pm$ 0.08 $\mu_B$/Fe (See Figure S2 of the SI file \cite{SI}). 
The relative cationic distribution, and reported in Table \ref{tab1}, have been determined in previous resonant elastic X-ray scattering (REXS) measurements performed on GFO thin films obtained from the same ceramic target \cite{lefevre2016determination} as this study. Oxygen K-edge XAS measurements as a function of the linear polarization were used to compare the electronic structure of our GFO films to stoichiometric GaFeO$_3$ single-crystals. The spectra show a strong polarization dependence reflecting a strong orbital in-plane/out-of-plane anisotropy. XAS at the O-K edge is related to the degree of covalency in the system, hence, reflecting the oxygen projected unoccupied density of states \cite{de1989oxygen}. In Figure \ref{fig2}a we show O-K edge XAS spectra acquired with linear horizontal (LH) and vertical (LV) polarized light with an incidence angle of $ca.$ 60$^\circ$ (grazing incidence (GI) geometry), with respect to the sample normal (see Inset of Figure \ref{fig2}a). In this geometry, the spectra acquired with LH polarization give mostly information regarding the unoccupied O$2p$ levels oriented out-of-plane ($i.e.$ parallel to the $\bf b$-axis of the film), while with LV we collect information of the unoccupied O$2p$ states oriented in the sample plane, $i.e.$ parallel to the $\bf a$- and the $\bf c$-axes, as both directions are essentially indistinguishable ($cf.$ Inset of Figure \ref{fig1}a). Indeed, LH and LV O-K edge spectra acquired in normal incidence (NI) geometry are identical (See Figure S3 of the SI file \cite{SI}), hence, confirming the GFO growth mode as sketched in the inset of Figure \ref{fig1}a. 
The high energy region, above 532 eV, is characterized by broad peaks due to the hybridization between O$2p$ and $4s$,$p$ Ga and Fe orbitals. On the other side, the pre-peak region, up to 532 eV, being related to the O$2p$ hybridization with Fe$3d$ orbitals is the one we analyzed. The two main features, labelled A- and B-, are roughly associated to the Fe$3d$ t$_{2g}$ and e$_{g}$-bands, which within the trigonal crystal field can be linked to (a$_1^g$, e$_g^{\pi}$) and e$_g^{\sigma}$ orbitals derived bands (Figure \ref{fig2}b). Thus, the peak A is a convolution between in-plane (e$_g^{\pi}$) and out-of-plane (a$_1^g$) derived bands. If the symmetry is further lowered, the e$_g^{\sigma}$ peak can also split. As matter of fact, the data acquired with LH and LV polarizations show a shift in the position of the A and B features which can be quantified by two-peaks Voigt profile (open symbols), shown as red dashed lines in Figure \ref{fig2}a. The analysis reveals that the average distance between A and B features, roughly corresponding to the 10Dq crystal field splitting, is of the order of 1.2 eV, while the energy splitting between a$_1^g$ and e$_g^{\pi}$ bands is of the order of 0.02 eV. Larger is the shift of the B-peaks ($\sim$0.1 eV), which demonstrates also a removal of the e$_g^{\sigma}$ degeneracy.  Furthermore, the strong intensity variation of the A and B peaks measured with LH and LV polarizations, suggests a strong anisotropic hybridization among Fe$3d$ and O$2p$ orbitals for both in-plane and out-of-plane directions. In Figure \ref{fig2}b we show a tentative energy diagram of the orbital hierarchy derived from the O K-edge XAS spectra analysis. In particular, it emerges that the out-of-plane (a$_1^g$) orbitals set higher in energy compared to in-plane ones (e$_g^{\pi}$), pointing at a trigonal elongation of the FeO$_6$ octahedra (See Figure S4 of the SI file \cite{SI}). Summarizing, our O-K edge measurements confirming the removal of the e$_g^{\sigma}$ orbitals degeneracy with an anisotropic Fe$3d$-O$2p$ hybridization path, point at a net non-zero $\mu_L$ value also for our GFO thin films, as already inferred for GaFeO$_3$ single-crystals \cite{kim2006orbital}. Here, the result of inner movements of Fe ions brings to a parity symmetry breaking of the $3d$ electron clouds with a subsequent apparition of a measurable magnetic orbital moment. Finally, it should be mentioned also that while O-K edge spectra cannot be used to precisely determine the values of the crystal field splitting parameters they can give, nevertheless, a good estimation.\\
In order to determine if the observed anisotropic Fe$3d$-O$2p$ hybridization network induces an orbital moment in our films, we performed XAS and XMCD measurements at the Fe L$_{3,2}$-edges in both NI and GI geometries. Figure \ref{fig3}a shows the area-normalized XAS spectra corrected from saturation effects \cite{nakajima1999electron} (more details are present in the SI file \cite{SI}). This correction is fundamental when a quantitative analysis of the magnetic orbital moment is required, and one of the most important parameters to consider is the X-ray penetration depth that depends upon the material density. 
In the case of our GFO thin films we calculated the X-ray penetration depth values from our XAS spectra as reported in the SI file \cite{SI}. As a result of this calculation we used in NI (GI) the values of 19.3 (24.5) nm and 83.2 (76.8) nm for the L$_3$ and L$_2$ edges, respectively. The overall acquired XAS signal well reproduces the features of the Fe L$_{3,2}$-edges of GaFeO$_{3}$ single crystals \cite{kim2006orbital} demonstrating the high-quality of our thin films. However, by a closer look to the XAS spectra (see inset of Figure \ref{fig3}a), in NI the relative intensities of the two peaks at L$_2$ is opposite to the spectrum acquired in GI, possibly reflecting the spatial anisotropy of the spin/orbital density on each of the Fe ions due to their relative displacements within the FeO$_6$ octahedra. \\
To get some basic information regarding the valence state of the Fe ions from the XAS spectra we resorted to atomic multiplet calculations by using the freely-available CTM4XAS software \cite{ctm4xas}. The simulations were performed neglecting the symmetry lowering provoked by the Fe displacements. While no substantial changes in the spectral line shape are expected from this approximation, it is strictly necessary to implement the off-centering of the Fe ions within the FeO$_6$ octahedra for a correct reproduction of the XMCD spectra and related integrals. Charge transfer (CT) mechanism was also implemented which, describing a dynamic charge fluctuation among the O2p and Fe3d bands, accounted for the large Fe$3d$-O$2p$ hybridization degree as demonstrated by O-K edge measurements. Simulations of the XAS spectra were implemented starting from the work of Chen $et$ $al.$ about magnetite thin films \cite{chen2004magnetic}. Crystal field (10D$_q$), CT ($\Delta$, U$_{dd}$, U$_{pd}$, T$_e$, T$_{ab}$) and relative occupations of Fe$^{3+}$ in both O$_h$ and T$_d$ sites are listed in Table \ref{tab2}. We considered a reduction of the Slater-Condon integrals of 70\%. The core-level spin-orbit coupling for the T$_d$ (O$_h$) simulation was set to 0.97 (1).
\begin{table}[h!]
\caption{\label{tab2} Optimized parameters with related weights for XAS spectra.}
\begin{ruledtabular}
\begin{tabular}{lccccccccc}
&10D$_q$&$\Delta$&U$_{dd}$&U$_{pd}$&T$_e$&T$_{ab}$&Weight   \\
& & & & & & &for XAS\\
Fe$^{3+}$(O$_h$)&1.2&3.0&6.0&7.5&0.5&1&92\%\\
Fe$^{3+}$(T$_d$)&1.2&3.0&6.0&7.5&0.5&1&8\%\\
 \end{tabular}
 \end{ruledtabular}
 \end{table}
\\
The spectral line features of the XAS data in GI are overall reproduced (red solid line of Figure \ref{fig3}a) accounting for Fe$^{3+}$ ions in both O$_{h}$ and T$_{d}$ sites with the expected proportion (see Tables \ref{tab1},\ref{tab2}). In NI we could not reproduce the exact L$_2$ doublet profile most likely because the relative intensities might depend on details of the crystal field. Still, we can exclude a different sites occupations, e.g. a larger amount (20\%) of Fe$^{3+}$ in Ga1 (T$_d$) sites (solid black line of Figure \ref{fig3}a), as-well-as the presence of a fraction of Fe$^{2+}$ in the system (orange line in Figure \ref{fig3}a). The overall good match between XAS data and simulation suggests that spurious effects, indeed, mainly attributable to a reduction of the Fe valence state and/or altered sites occupations, can be neglected for the following discussions. \\
Figures \ref{fig3}b,c report the XMCD spectra acquired for both NI and GI geometries after the saturation correction. Also in this case we can highlight a good agreement with data reported for GaFeO$_{3}$ single-crystals showing similar features, alike the two negative peaks at L$_3$ and a positive feature at L$_2$ (Figure S5 of the SI file \cite{SI} shows the two XMCD spectra superimposed). By focusing on the integral sign of the XMCD, while in GI we recover a negative value (corroborating the reported results for bulk GaFeO$_{3}$), in NI we surprisingly observe a positive integral value. To guarantee a quantitative analysis of this finding any drawbacks linked to the practical application of the sum-rules need to be excluded. Indeed, the extrapolation of absolute values for orbital and spin moments is particularly challenging \cite{Elnaggar2020}, especially in the case of materials containing the magnetic ions in the same nominal oxidation state but in different coordination and symmetry \cite{wu1993first} ($i.e.$, mixed occupation of T$_d$  and O$_h$  sites), as it is for GFO. XMCD measurements performed on undoped GFO thin films, for which the Fe$^{3+}$ occupation of T$_d$ sites is practically zero \cite{lefevre2016determination}, rendered the same integral sign variation as for doped GFO thin films (See Figure S1 of the SI file for more details \cite{SI}). Importantly, we obtain the result that sum-rules can formally hold also for our doped GFO thin films. In this framework, the integral of the XMCD is proportional to the average orbital moment \cite{Chen1995, thole1992x}, and we get the unexpected result that the relative orientation of the calculated average orbital and spin moments depends upon the crystallographic direction. Table \ref{tabXMCD} reports the average $\mu_L$ and $\mu_S$ values, as obtained from sum-rules, for both NI and GI geometries. Within the systematic error that accompanies the application of the sum-rules of about 10-20\%, we found that the external magnetic field is strong enough to align $\mu_S$ along both crystallographic directions, but this does not hold for $\mu_L$ which, within the error, has values that do not overlap. As already discussed by Kim $et$ $al.$, intrinsic Fe-displacements within the most distorted O$_h$ sites of the GFO orthorhombic crystal structure are responsible of a non-zero $\mu_L$. The latter extending perpendicular to the off-centering direction is a result of the anisotropic electron cloud around the Fe ions\cite{kim2006orbital}. In this picture $\mu_L$ is linked to the magnitude/direction of the Fe off-centering with respect to the FeO$_6$ center of gravity, and to fully orient it along H, one should assume that the Fe-displacements can take place along all the crystallographic directions within the distorted FeO$_6$ octahedra. As a matter of fact, our XMCD data ($cf.$ Figure \ref{fig3}c), verified on several samples and showing a sign-reversal of the XMCD integrals, tell us that this is not the case and, on the contrary, inform about a possible large lattice-orbital effect characterizing the most distorted Fe1 and Fe2 sites of our GFO thin films.\begin{table}[h!]
\caption{\label{tabXMCD} XMCD sum-rules-derived $\mu_L$ and $\mu_S$ for NI and GI geometries at 6.5 T. Number of holes in the 3d states equal to 4.8 to account for the CT.}
\begin{ruledtabular}
\begin{tabular}{lccccccccc}
&$\mu_S$  & $\mu_L$  & $\mu_{total}$    \\
&$[\mu_B/Fe$]&[$\mu_B/Fe]$&[$\mu_B/Fe$]\\
NI&+0.72 $\pm$ 0.07 &-0.023 $\pm$ 0.002&0.70 $\pm$ 0.07\\
GI&+0.68 $\pm$ 0.07&+0.016 $\pm$ 0.002&0.70 $\pm$ 0.07\\
 \end{tabular}
 \end{ruledtabular}
 \end{table}

\begin{figure}
\includegraphics[width=0.5\textwidth, keepaspectratio]{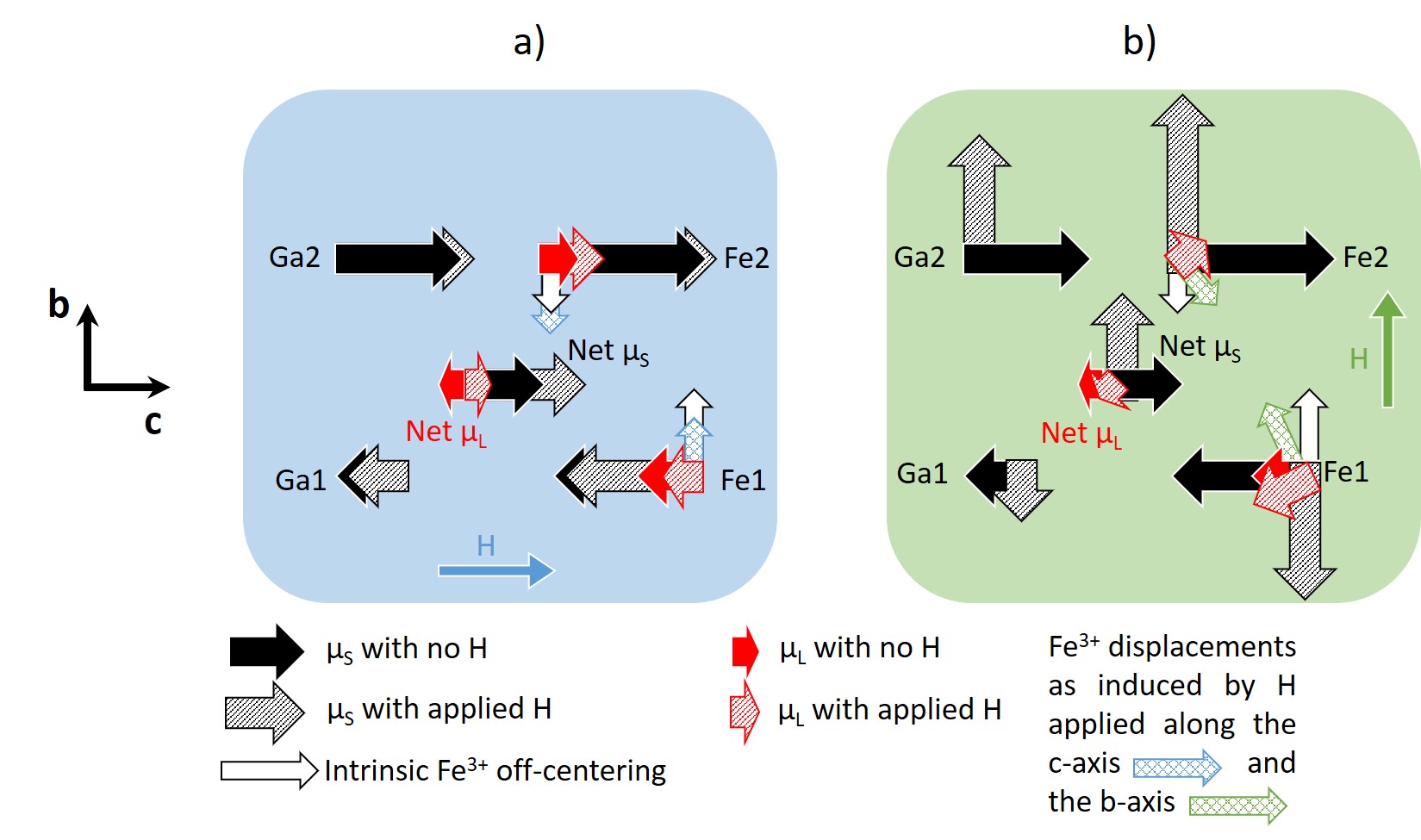}
\caption{\label{fig5} a) Parallel configuration among $\mu_L$ and $\mu_S$ and (b) their geometrical construction under magnetic field along the $\bf b$-axis for the anti-parallel configuration. The length of the arrow representing  the spin moment  is proportional to the Fe$^{3+}$ occupancy at each site.}
\end{figure}

As discussed above, the off-centering of the Fe$^{3+}$ ions determines a different degree of hybridization between Fe$3d$ and O$2p$ states along the main crystallographic axes for each distorted sites, hence, establishing anisotropic Fe$3d$-O$2p$ hybridization paths. As already shown by Arima $et$ $al.$, Fe$^{3+}$ ions sitting in the most distorted Fe1 and Fe2 sites move along the $\bf b$-axis in opposite directions \cite{Arima2004a} (plain white arrows sketched in Figure \ref{fig5}). In the absence of H such intrinsic off-centering sets a net $\mu_L$ along the $\bf c$-axis, opposed to the net $\mu_S$ for which the contribution from Fe2 and Ga2 sites prevails over the one of Fe1 and Ga1 sites, the latter being mostly occupied by Ga ions. When we measure the XMCD in GI geometry, being H mostly oriented in the sample plane, the Fe ions re-adjust their position in the perpendicular direction \cite{Arima2004a}, as exemplified by the blue shaded arrows of Figure \ref{fig5}a. As a direct consequence of this, the orbital moment weakens (expands) at Fe1 (Fe2) site, hence, promoting a net $\mu_L$ parallel to the net $\mu_S$ and confirming our experimental finding in GI geometry. On the other side, in NI geometry, H is directed out-of-plane along the $\bf b$-axis which represents the magnetic hard axis and, at high enough field (6.5 T), the spin structure can be modified as sketched in Figure \ref{fig5}b (See Figure S2 of the SI file \cite{SI}). One then expects the orbital magnetic moment to follow this rotation. This means that, since $\mu_L$ is largely controlled by the anisotropic displacements of the Fe$^{3+}$ ions within the distorted FeO$_6$ octahedra at Fe1 and Fe2 sites, the Fe$^{3+}$ off-centering should go mainly along the $\bf c$-axis foreseeing a complete re-orientation of the distortions within the GFO crystal structure. Without considering the energy scale associated to this process, if this happens, the sign between $\mu_L$  and $\mu_S$ would not change, since we will meet the same situation as introduced and explained for GI geometry (Figure \ref{fig5}a), with no possibility to distinguish among the $\bf b$- and $\bf c$-axes anymore. On the contrary, if we consider that the Fe$^{3+}$ ions displacements have a large component along the $\bf b$-axis, which implies that $\mu_L$ at each Fe1 and Fe2 site only slightly rotates, as sketched in Figure \ref{fig5}b by the green shaded arrows, the net $\mu_L$ along H is now oriented antiparallel to $\mu_S$ as, indeed, expected from our XMCD results in NI geometry (Figure \ref{fig3}c).\\
This scenario explains the anomalous change of the average orbital moment with the crystallographic direction and has important implication for the physics of this system. In particular, it shows that a strongly correlated oxide may have a direction-dependent average magnetic orbital moment as a result of anisotropic distortions and related lattice-orbital effects, which geometrically control the relative orientation between the spin and orbital moments. Thus, a control over the distortions at each site of the GFO crystal structure would then offer an adjustment knob over the spin-orbit coupling in the system. Furthermore, Elnaggar $et$ $al.$ showed that a non-collinearity between spin and orbital magnetic moments is possible also in magnetite (Fe$_3$O$_4$) single crystals \cite{Elnaggar2020}. There, an intimate interplay between trigonal distortions, spin-orbit coupling and exchange interaction were assumed to be at the origin of the non-collinear orbital and spin ordering direction. In GFO thin films, this non-collinearity is mainly linked to the intrinsic Fe off-centering within the distorted FeO$_6$ octahedra with direct consequences on the magnetoelectric and magnetic properties. We believe that our results will encourage the study of TMOs materials exhibiting multiferroic properties to assess their potential in spintronics by performing experiments, such as spin-pumping, ferromagnetic resonance and/or spin-Seebeck effects, taking into account the relative orientation between H and crystallographic axes.

\begin{acknowledgments}
This work was funded by the French National Research Agency (ANR) through the ANR MISSION ANR-18-CE-CE24-0008-01. DP has benefited support from the initiative of excellence IDEX-Unistra (ANR-10-IDEX-0002-02) from the French national program Investment for the future. S.H. acknowledges financial support from ANR through Grant ANR-17-EURE-0024 EUR QMat. The synchrotron experiments were performed at SOLEIL synchrotron facility in France under proposal number 20180460. The authors wish to thank F. Choueikani for the scientific and technical support at DEIMOS and L. Joly, J.P. Kappler for preliminary measurements performed at SLS. P. Ohresser, M. Alouani, and P. Sainctavit are also deeply acknowledged for fruitful and inspiring discussions regarding the possible mechanism explaining our XMCD data. F. de Groot is acknowledged for advice regarding the atomic multiplet calculations. Finally, the authors are very grateful to S. Grenier for the implementation of the DYNA code used for the calculation of X-ray penetration depths.
\end{acknowledgments}

\bibliographystyle{nature}
\bibliography{gfo}
\end{document}